\documentstyle[epsfig]{aipproc}

\newcommand{\mej}{M_{\rm ej}}
\newcommand{\mc}{M_{\rm cir*}}
\newcommand{\msw}{M_{\rm sw}}
\newcommand{\mswi}{M_{\rm sw, is}}
\newcommand{\ein}{E_{\rm in}}
\newcommand{\beq}{\begin{equation}}
\newcommand{\eeq}{\end{equation}}
\newcommand{\lrho}{{\ell_\rho}}
\newcommand{\vej}{v_{\rm ej}}
\newcommand{\caln}{{\cal N}}
\newcommand{\nsh}{\caln_{*h}}

\begin{document}
\title{Young Supernova Remnants: Issues and Prospects}

\author{Christopher F. McKee}
\address{University of California, Berkeley, CA USA}

\maketitle

\begin{abstract}

	The dynamical evolution of young supernova remnants (YSNRs) is
governed by the density distribution in the ejecta and in
the ambient medium.  Analytic solutions are available for
spherically symmetric expansion, including the transition
from the ejecta-dominated stage to the
Sedov-Taylor stage.  YSNRs serve as valuable physics laboratories,
in which we can study nucleosynthesis, the early evolution of compact
objects, pulsar physics,
particle acceleration, the formation and destruction
of dust, hydrodynamics at high Reynolds numbers, shock physics
at high Mach numbers, and the effects of thermal conduction in
interstellar plasmas.
There are several challenges in YSNR research:
(1) Where are the very young remnants in the Galaxy?  We expect 5-10 
to have occurred since
Cas A, but with the possible exception of a remnant reported at 
this conference, none have been seen.  (2) Can very young SNRs produce
gamma-ray bursts?  The acceleration of a shock in the outer
layers of a supernova, first suggested by Colgate, can account
for gamma-ray bursts such as that believed to be associated
with SN 1998bw, and more powerful explosions can account for
the energies seen in many cosmological bursts.  (3) The Connections Challenge:
Can one infer the nature of the supernova and its progenitor star
from observations of the YSNR?

\end{abstract}

\section*{Introduction}

	Young supernova remnants (YSNRs) are fascinating objects.
Just as star formation connects the interstellar medium (ISM) with
stars through the process of gravitational contraction, so YSNRs
complete the connection back to the ISM through a titanic explosion driven by
the release of either gravitational or nuclear energy.  
An SNR is produced by the interaction of the ejecta with the ambient
medium or, alternatively, by the effects of a pulsar left behind
by the explosion.
The ambient medium can be circumstellar---i.e.,
matter ejected by the progenitor or its companion---or
interstellar.  I shall use the term ``YSNR'' to denote
an SNR in which the 
total mass ejected by the star, both the circumstellar mass, $\mc$,
and the mass ejected by the supernova explosion, $\mej$, 
exceeds the mass of swept-up interstellar
gas, $\mswi$---i.e., $\mc+\mej>\mswi$.  

	In this
article, I shall first review the dynamics of YSNRs.  I shall then
discuss YSNRs as physics laboratories that enable us to address
problems that are difficult or impossible to address on Earth.
Finally, I shall address several challenges posed by YSNRs, including
whether they can be the progenitors of gamma-ray bursts.

\section*{Review of YSNR dynamics}

	The dynamics of a YSNR is driven by the interaction between
the ejecta and the surrounding matter.  The first step in analyzing
the dynamics is therefore to determine the density structure of
the ejecta.  Chevalier \cite{chev76} began this effort by
working out the dynamics of the blast
wave produced by a core-collapse supernova as it propagated in
the envelope of a red giant.  Subsequently, Chevalier \& Soker
\cite{chev89} approximated the envelope of a blue supergiant
as having a power-law density distribution with $\rho_0\propto
r^{-17/7}$, which gives rise to a Primakoff blast wave in which
the density and velocity are power-laws in radius.  A general
treatment of the generation of the ejecta density profile 
(under the assumption of spherical symmetry) was
given by Matzner \& McKee \cite{mm99}, and we shall briefly 
review their results here.

	They began by developing an analytic expression for
the velocity of the supernova shock as it propagates through the
progenitor star.  In the interior, one expects $v_s\propto [\ein/m(r)]
^{1/2}$, where $m(r)\equiv M(r)-M_{\rm rem}$ is the mass of the 
ejecta inside $r$.  When the shock reaches the atmosphere, it
accelerates down the density gradient according to
$v_s\propto [m(r)/\rho_0(r)r^3]^\alpha$, with $\alpha\simeq 0.19$.
A general expression for the shock velocity, that is accurate
throughout the envelope and atmosphere, is \cite{mm99}
\beq
v_s=A\left[\frac{\ein}{m(r)}\right]^{1/2}\left[\frac{m(r)}{\rho_0
	(r)r^3}\right]^{\alpha}.
\label{eq:vs}
\eeq
It is possible to evaluate the coefficient $A$ analytically as well
\cite{tan00}.  This expression is accurate to within 2\% in
stellar envelopes.  It should be contrasted with the result of
the Kompaneets approximation, which gives $v_s\propto
[\ein/\rho_0(r)r^3]^{1/2}$ and fails in both the interior and the
atmosphere.  

	With this result as a base, Matzner \& McKee were able
to determine approximate analytic expressions for the density
distribution of the ejecta that are quite general.  The distribution
in the outer ejecta can be approximated as a power-law
in velocity,
\beq
\rho\propto (\vej/v)^\lrho,
\eeq
where $\vej$ is the maximum velocity of the ejecta.
In general, $\lrho$ can depend on position; indeed, by
comparing with numerical models,
Dwarkadas and Chevalier \cite{dwar98} find that an
exponential approximation fits best for
Type Ia SNRs.  In the outer ejecta, the value of $\lrho$ is 
typically greater than 5, which means that the energy as
well as the mass of the ejecta are concentrated in the interior.

	In the ejecta-dominated stage of evolution of
SNRs ($\mej>\msw$), there are two shocks: the blast-wave shock
that advances into the ambient medium at velocity $v_b$ and
a reverse shock that propagates back into the ejecta with
a relative velocity $v_r$ \cite{mck74}.  Between the shocks lie
the shocked ambient medium and the shocked ejecta, separated by
a contact discontinuity.  For the case of a steep density gradient
in the ejecta ($\lrho>5$), the ejecta-dominated stage of evolution
can be divided into two parts:  First, there
is a brief
initial stage described by the Hamilton-Sarazin \cite{ham84} similarity
solution, in which the velocity of the blast-wave shock $v_b$ is
approximately equal to $\vej$.  In a medium of constant
density, the radius of the blast wave expands with time as
$R_b\simeq 1.10\vej t(1-at^{3/2})$, where $a$ is a numerical
constant.  The velocity of the reverse shock propagating
back into the ejecta satisfies $v_r\propto t^{3/2}$.
Once the blast-wave velocity has slowed significantly below
$\vej$, the evolution enters a second self-similar phase of 
evolution, which is described by the Chevalier-Nadyozhin 
solution \cite{chev82} \cite{nad85}.  In this solution,
the blast-wave radius varies as $R_b\propto t^{1-3/\lrho}$;
the velocities of the blast-wave shock and the reverse shock
both scale as $t^{-3/\lrho}$.

	When the mass of the swept-up ambient medium exceeds
the mass of the ejecta, 
the dynamics approaches that
of an adiabatic blast wave produced by a point explosion.
The Sedov-Taylor similarity solution for this problem has
$R_b\propto (\ein t^2/\rho_0)^{1/5}$.  Many historical remnants
are in transition between the ejecta-dominated stage and the 
Sedov-Taylor stage, and McKee \& Truelove \cite{mck95}
proposed an approximate
analytic solution that describes the evolution from the 
Chevalier-Nadyozhin stage through the Sedov-Taylor stage.
This solution was developed by Truelove \& McKee
\cite{true99} (note the erratum in \cite{true00}), who
defined the characteristic quantities
\begin{eqnarray}
R_{\rm ch}&\equiv&(\mej/\rho_0)^{1/3},\\
v_{\rm ch}&\equiv&(\ein/\mej)^{1/2},\\
t_{\rm ch}&\equiv& R_{\rm ch}/v_{\rm ch},
\end{eqnarray}
and then
gave approximate expressions for $R_b/R_{\rm ch}$, $v_b/v_{\rm ch}$,
and $v_r/v_{\rm ch}$ as functions of $t/t_{\rm ch}$
(or in some cases, $t/t_{\rm ch}$ in terms of $R_b/R_{\rm ch}$, 
etc).  This solution
has been successfully compared with the observations of Tycho's SNR
by Hughes \cite{hug00}.

	Actual YSNRs present a rich range of phenomena that go
well beyond the simple  dynamics described
above.  For example, if the YSNR has an embedded pulsar, then
the pulsar nebula can be dramatically transformed by the reverse
shock \cite{chev98}.  Deviations from spherical symmetry can have
major effects.  Such deviations can be due to asymmetries in
the explosion (e.g., \cite{blon00}), the ambient medium \cite{maj98},
or due to instabilities (e.g., \cite{gull73} \cite{chev92}).
Theoretical and observational studies of these effects are given
in these conference proceedings.

\section*{YSNRs as Physics Laboratories}

	Young SNRs provide extreme environments in which novel
physical effects can be studied.  The role of YSNRs in cosmic
ray acceleration is reviewed elsewhere in this volume by
Steve Reynolds.  Some YSNRs harbor pulsars, and these objects
are discussed by Manchester and by Gaensler elsewhere
in these proceedings.  Here I shall
briefly comment on how YSNRs can serve as physics laboratories
for the study of nucleosynthesis, dust formation and destruction,
and interstellar hydrodynamics.

\subsection*{Nucleosynthesis}

	YSNRs provide laboratories to test both
the theory of the formation of the elements and the theory of
supernova explosions.  X-ray spectroscopy is crucial, since
optical spectroscopy is often sensitive to only a small fraction
of the mass of the ejecta.  Spatially resolved X-ray spectroscopy
is now available with {\it Chandra} and with {\it XMM-Newton}.  The
Astronomy and Astrophysics Survey Committee (AASC) \cite{aasc} has recommended
that an even more powerful instrument be built during the coming
decade, {\it Constellation-X}.  Consisting of four X-ray telescopes,
{\it Constellation-X} would have an energy resolution $E/\Delta E\sim
300-5000$ over the energy range 0.25-40 keV.  Its effective area would
be about 20-100 times that of existing instruments, and it would
have an angular resolution of about 15 arcsec.
A particularly strong clue on the nucleosynthesis that occurs in YSNRs
is provided by radioactive elements, which can be studied with
the INTEGAL spacecraft that is due to be launched in 2001.  The
Panel on High-Energy Astrophysics from Space of the AASC recommended
an Explorer Class mission for nuclear line X-ray spectroscopy.

\subsection*{Dust Formation and Destruction}

	Most of the refractory elements like silicon and iron in
the interstellar medium are contained in dust grains.  Observations
of isotopic anomalies in meteorites suggest that dust forms in
core-collapse supernovae \cite{clay75}.  This idea received
observational confirmation when SN1987A showed several signs of
dust formation, including dust emission, dust absorption,
and a drop in the line intensities of the refractory
elements at the time at which the dust emission appeared 
\cite{dwek98}.  
Whereas observations of dust emission suggest that only about
$(1-10)\times 10^{-4} M_\odot$ of dust formed in the supernova, 
observations of the extinction
and of the diminution of the refractory lines---which measure all
the dust, not just the hot dust---suggest that most of 
the refractory elements in the ejecta could have gone into dust
\cite{dwek98}.   From a
theoretical perspective, supernovae have long been thought to
be sites of dust formation: Once grains are injected
into the ISM they are subject to efficient destruction by
SNR shocks, so it is essential that the refractory
elements be injected in solid form in order to
account for their large depletions (e.g., \cite{ds79}).
Even in this case, significant growth of refractory
grains is inferred to
occur in the ISM  \cite{dwek98} \cite{mck89}.  

	However, the same shock processes
that are effective at destroying grains in the ISM can destroy them
in YSNRs.  How can the grains survive the reverse shock?  How
can they survive being embedded in the hot gas in the interior
of an SNR?  How can they survive being decelerated from velocities
in excess of $10^3$ km s$^{-1}$?  Sputtering is the dominant
process of grain destruction in this case \cite{tiel94} \cite{dwek96}.
The effectiveness of sputtering is enhanced by the shattering of 
the grains that occurs in shocks, since that increases the
grain surface area \cite{jone96}.  Furthermore, recent ISO observations have
cast doubt on the idea that core collapse supernovae are the dominant 
source of interstellar grains: Douvion, Lagage, \& Cesarsky
\cite{douv99} argue that only a small fraction of the silicon
in Cas A is microscopically mixed into the regions necessary to make silicates,
whereas Arendt, Dwek, \& Mosely \cite{aren99} find that the spectrum
of Cas A indicates that the silicates that do form are not
of the type that is typical of interstellar grains.  

	In order to understand how dust is formed and destroyed
in YSNRs, further observational and theoretical work is needed.
The {\it Space Infrared Telescope Facility (SIRTF)} and the 
{\it Stratospheric Observatory for Infrared Astronomy (SOFIA)} will
provide valuable new data.  Further in the future, the 
{\it Next Generation Space Telescope (NGST)} offers the possibility
of observing the infrared spectra of grain-producing elements and
of the grains themselves at high angular resolution.

\subsection*{Interstellar Hydrodynamics}

	YSNRs offer a fascinating laboratory in which to study
hydrodyamic processes in the ISM.  
As described by McCray elsewhere in these proceedings,
observations of the youngest
nearby SNR, the remnant of SN1987A, even provide us with an 
opportunity to observe
the {\it temporal evolution} of these hydrodynamic
processes.  

	YSNRs produce extremely powerful
shocks, with shock velocities that can exceed $10^4$ km s$^{-1}$ and
Mach numbers $\cal M$ that can exceed $10^3$.  These shocks may be 
strongly modified by the acceleration of cosmic rays.  Indeed,
Hughes, Rakowski, and Decourchelle \cite{hug00b} have analyzed
{\it Chandra} observations of the SMC SNR E0102.2-7219 and have
concluded that the most natural interpretation of the low electron
temperatures they observe is that most of the shock energy has gone
into cosmic rays.  A key issue of shock physics that can be addressed
through observations of YSNRs is the degree of collisionless heating
of electrons behind fast shocks.  Ghavamian et al \cite{ghav01} have
studied optical emission from nonradiative shocks in several SNRs
(the Cygnus Loop, RCW 86, and Tycho), and they infer that the
electron heating efficiency falls as the shock velocity increases: they find
$T_e/T_i\sim 0.7-1$ in the Cygnus Loop ($v_s\sim 300$ km s$^{-1}$),
$\sim 0.3$ in RCW 86 ($v_s\sim 600$ km s$^{-1}$),
and $\lesssim 0.1$ in Tycho ($v_s\sim 2000$ km s$^{-1}$).  However, they did
not allow for the possibility that some of the energy of the shocked
gas is in cosmic rays, as found by Hughes et al \cite{hug00b}.
Further studies of shock physics using both optical
and X-ray observations will be very illuminating.

    	A problem of fundamental importance in astrophysics is
the interaction of a shock wave with a density inhomogeneity
(a ``cloud'').
Astrophysical plasmas are generally inhomogeneous,
and shocks are ubiquitous because these plasmas can typically cool
to temperatures well below those associated with the violent
events that occur in them.  Once the shock has passed the cloud,
this problem reduces to that of a cloud embedded in a flow,
or, equivalently, a clump of ejecta interacting with the ambient
medium (an ``interstellar bullet'').
These problems have been studied both theoretically
(e.g., \cite{mck75}, \cite{mck88}, \cite{kle94}) and in
the laboratory \cite{kle00}.  The clouds are subject to both
thermal evaporation and hydrodynamic stripping; the first process
plays a key role in
the three-phase model of the ISM \cite{mck77}.  Magnetic
fields have a strong effect on both processes; 
the effective conductivity in a magnetized, collisionless
plasma remains uncertain.  The combined effects of hydrodynamic
stripping and thermal conduction are invoked in the model
of turbulent mixing layers \cite{beg90} \cite{sla93}, in which
it is assumed that conduction is so efficient
that the mixed gas starts with an initial temperature
that is determined only by mass and energy conservation, but
not so efficient that it affects the radiative cooling of 
the gas.  As yet, there is no direct observational evidence
that the boundary layers of shocked clouds are described
by turbulent mixing layers, but
YSNRs provide an ideal laboratory for studying this question,
as well as for determining the fate of the ``interstellar bullets'' seen
in the jet in Cas A \cite{fes96} and for addressing many other
problems in interstellar gas dynamics.

\section*{Challenge: Where are the Very Young SNRs in the Galaxy?}

	No supernova remnants are known in the Galaxy since
Cas A exploded more than 300 years ago, with the possible
exception of a very young SNR in the Galactic Center reported
at this meeting \cite{sen01}.  Van den Bergh \& Tammann
\cite{van91} cite a range of estimates for the
Galactic supernova rate; perhaps the best is from observations
of SNe in other galaxies, which gives $8.4h^2$ per century,
where $h$ is the Hubble constant normalized to 100 km s$^{-1}$
Mpc$^{-1}$.  Taking $h=0.63$ from a recent ``concordance model'' based 
on observations of the cosmic microwave background and large scale
structure \cite{teg00}, we find a Galactic SN rate of 3.3 per century.  
Subsequently, van den Bergh \& McClure \cite{van94} estimated
a rate of $1.7-1.9$ per century for the same value of $h$.
McKee \& Williams \cite{mck97} analyzed data on the OB associations
in the Galaxy and estimated that the rate of core-collapse
supernovae is 2.0 per century; including Type Ia's, the total
SN rate becomes 2.3 per century (note that this estimate is
independent of $h$).  In 300 years, we therefore expect that 
5-10 SNe have exploded in the Galaxy.  Where are their remnants?

	Most of these SNe are core-collapse SNe that originated
in associations.  McKee \& Williams \cite{mck97} worked out
the birthrate of associations in the Galaxy based on observations
of radio H~II regions, radio free-free emission, and N~II
$\lambda 122~\mu$m emission.  
They adopted an IMF based on the work of Scalo \cite{sca86},
in which $d\caln_*/d\ln m_*\propto m_*^{-1.5}$, slightly steeper
that the Salpeter value.
If $8 M_\odot$ is the lower mass limit for core-collapse SNe,
then half the missing SNRs had progenitors less massive than 13 $M_\odot$.
Under the assumption that each
association has five consecutive generations of star formation
(lasting for a total of 18.5 Myr), 
they estimated that the birthrate
of associations that eventually produce at
least $\nsh$ stars more massive than 8 $M_\odot$ is
\beq
\dot \caln_a(>\nsh)=0.33\left(\frac{7200}{\nsh}-1\right)~~~~
	{\rm Myr}^{-1},
\eeq
where the largest association in the Galaxy has $\nsh\simeq 7200$
high-mass stars.  They extrapolated this distribution down
to associations with 100 stars per generation, or 500 stars
altogether; for their
IMF, this corresponds to a minimum value of $\nsh\simeq 1.3$.
From this distribution, one can infer that half of
these missing SNRs occurred in associations in which about
35 SNe have already occurred.  To find out where in the Galaxy
the SNe are likely
to have occurred, we use the spatial distribution of associations
from reference\cite{mck97}
\beq
\frac{d \dot\caln_a}{dA}=\dot\caln_a(>\nsh)\left[\frac{\exp(-R/H_R)}
	{A_{\rm eff}}\right]~~~~~({\rm 3~ kpc}\lesssim R\lesssim 11~
	{\rm kpc}),
\eeq
with a radial scalelength $H_R\simeq 3.5$ kpc and an effective
area $A_{\rm eff}=47$ kpc$^2$.  From this we can infer that half
the missing SNRs are between 3 and 6 kpc from the Galactic
Center, which includes the highly obscured molecular ring; other
SNRs could lie on the far side of the ring, where the
obscuration is even greater.

	We conclude that many of the missing very young SNRs are either highly
obscured or located in regions that have been evacuated by 
previous SNRs.  In this latter case, the SNR is ``muffled'' by
the cavity: the ejecta remain in free expansion out to 
a radius of about $20[(M_{\rm ej}/10M_\odot)(10^{-2} {\rm cm}^{-3}
/n_{\rm H})]^{1/3}$ pc and radiate their energy less efficiently
than in denser environments \cite{mck88}.  YSNRs like Cas A, which
is interacting with circumstellar matter, or the Crab, which is
a plerion, would be easily seen throughout the Galaxy, however;
we conclude that the progenitors of such SNRs are relatively rare.

   	Two projects recommended by the AASC could dramatically
improve the census of YSNRs in the Galaxy:  The {\it Energetic X-ray
Imaging Survey Telescope (EXIST)} will survey the sky in the energy
range 5-300 keV with 300 arcsecond resolution.  It will be able
to detect highly obscured SNRs.  The {\it LOw Frequency ARray (LOFAR)} 
is a radio telescope with
a square kilometer of collecting area at wavelengths between
200 and 1000 cm and an angular resolution of 1 arcsec.  With
these capabilities, it will be extremely sensitive to the nonthermal
radio emission from YSNRs.

\section*{Challenge: Can Very Young SNRs Make Gamma-Ray Bursts?}

	Gamma-ray bursts (GRBs) remain one of the great puzzles in
astrophysics more than 30 years after their discovery (see
Piran \cite{pir99} for a review).  There are several arguments suggesting
that GRBs are associated with the deaths of massive stars: 
supernova explosions are the most energetic phenomenon known to occur
outside galactic nuclei; observations of GRBs with afterglows have
suggested that they may be associated with star-forming regions
(e.g., \cite{kul00}); and
finally, GRB 980425 appears to be temporally and spatially associated
with the unusual supernova SN1998bw \cite{gal98}.
Long before there was any observational evidence for an
association of GRBs with massive stars, Colgate \cite{col74} suggested
that GRBs could be produced by the emission from accelerating
shock waves in the outer layers of supernovae in galaxies
at distances of 30-50 Mpc.  It is now known that most GRBs are at
cosmological distances, and Paczynski \cite{pac98} has suggested
that more energetic stellar explosions associated
with the collapse of rapidly rotating massive stars (``hypernovae'') may
underlie GRBs.  Some attempts to model the energetics of a gamma-ray
burst produced by a spherical explosion of SN1998bw found that the
energy in relativistic ejecta was inadequate (e.g., \cite{woo99}).
Extrapolating from a nonrelativistic analysis, however,
Matzner \& McKee \cite{mm99} concluded that there was
enough energy in relativistic ejecta to produce the observed
burst.  What was missing from these analyses was the inclusion
of the relativistic effects on the hydrodynamics as the
shock wave accelerates to velocities near the speed of light.
This omission has been rectified by Tan, Matzner, \& McKee's 
analysis of trans-relativistic blastwaves in supernovae \cite{tan00}.

	Shock-acceleration models of GRBs have a great advantage
in that they naturally deal with the ``baryon-loading problem.''
The observed energies of GRBs are so great that any viable model
must be highly efficient, putting a significant fraction of the energy
into radiation.  Shock acceleration models 
are intrinsically efficient because they concentrate the energy
in the outermost, fastest moving material in an explosion.

	 The dynamics of a nonrelativistic shock in a stellar
envelope are described by equation (\ref{eq:vs}).  In the
outer layers of the envelope, where the enclosed mass
$m(r)$ approaches the total ejected mass $\mej$, this reduces
to
\beq
v_s\propto (\rho_0 r^3)^{-\alpha},
\eeq
where $\alpha\simeq 0.19$.  The shock accelerates if the stellar
density $\rho_0$ falls off faster than $1/r^3$.  Gnatyk \cite{gna85}
proposed that a similar relation should hold for trans-relativistic
flows, with $v_s$ replaced by $\Gamma_s\beta_s$ and $\alpha$ set equal
to 0.2 for accelerating shocks; here $\beta_s=v_s/c$ and
$\Gamma_s=(1-\beta_s^2)^{-1/2}$.  Tan et al \cite{tan00} obtained
a more accurate representation of the shock dynamics, which
in addition is valid throughout the star: 
\begin{eqnarray}
\Gamma_s\beta_s  & = & p(1+p^2)^{0.12}\\
p & \equiv & A\left[\frac{\ein}{m(r)}\right]^{1/2}\left[\frac{m(r)}
	{\rho_0 r^3}\right]^{0.187}.
\end{eqnarray}
This reduces to equation (\ref{eq:vs}) in the nonrelativistic
limit.

	After the passage of the shock through the stellar envelope,
the shocked gas is left with an internal energy that is equal
to its kinetic energy.  As the shocked gas undergoes adiabatic
expansion, this internal
energy is converted to kinetic energy; in addition, the inner
layers do work on the outer layers.  Matzner \& McKee \cite{mm99}
found that as a result the final velocity in a nonrelativistic,
planar flow is about
twice the post-shock value.  In the relativistic case, Tan et al
\cite{tan00} find
\beq
\frac{\Gamma_f\beta_f}{\Gamma_s\beta_s}=2.0+(\Gamma_s\beta_s)^
{\surd 3},
\eeq
where the scaling in the relativistic limit is the same as that
found analytically by Johnson \& McKee \cite{joh71}.  The effects
of spherical expansion reduce $\Gamma_f\beta_f$ somewhat below
the planar value.

    Based on their nonrelativistic theory, Matzner \& McKee
\cite{mm99} estimated the mass of relativistic
ejecta in a supernova; for example, for an $n=4$ polytrope, the
mass with $\Gamma_f\beta_f>1$ is
\beq
M_{\rm rel}=2.7\times 10^{-6}\left(\frac{\ein}{10^{52}~{\rm erg}}
       \right)^{3.29}\left(\frac{1~M_\odot}{\mej}\right)^{2.29}
       ~~~~~M_\odot.
\eeq
The corresponding energy is
\beq
M_{\rm rel} c^2=3.1\times 10^{48}\left(\frac{\ein}{10^{52}~{\rm erg}}
       \right)^{3.29}\left(\frac{1~M_\odot}{\mej}\right)^{2.29}
       ~~~~~{\rm erg}.
\eeq
Tan et al \cite{tan00} show that this is a reasonably good
estimate for $M_{\rm rel}$, and furthermore, that the 
kinetic energy of the ejecta
above some value of $\Gamma_f\beta_f$ is 
\beq
E_k\propto F(\Gamma_f\beta_f)M_{\rm rel}c^2,
\eeq
where (for $n=4$) $F\propto(\Gamma_f\beta_f)^{-4.69}$ in the nonrelativistic
limit and $F\propto(\Gamma_f\beta_f)^{-0.98}$ in the
relativistic limit.

	To determine if the explosion of SN1998bw had enough energy
to power GRB980425, Tan et al used a model of the supernova
kindly provided by Stan Woosley.  This model, which
is consistent with the observed light curve of the supernova,
had a Wolf-Rayet progenitor with a 
mass of $6.55 M_\odot$, an ejecta mass of $4.77 M_\odot$,
and an energy of $2.8\times 10^{52}$ erg.  Both the analytic theory
and numerical simulation yield about $2\times 10^{48}$ erg in
ejecta with $\Gamma_f\beta_f>1$; by comparison, the observed
energy of the burst was $(8.1\pm 1.0)\times 10^{47}$ erg,
under the assumption that it was in fact associated with the
supernova \cite{gal98}.  The energy content of the ejecta
therefore appears sufficient to power the burst.  In order
to release the energy, however, it must first be randomized.
This can be done through interaction with the dense stellar
wind expected for a Wolf-Rayet star \cite{woo99}; a mass loss rate of
a few times $10^{-4} M_\odot$ yr$^{-1}$ is required.
Because the emission is due to the interaction with
the ambient medium, this model for GRBs is in fact
a YSNR model.
Tan et al did not attempt to carry out a calculation of
the emitted spectrum, so it remains to be seen whether an
approximately spherical shock-acceleration model can account
for the observed burst.  However, they did show that the
characteristic synchrotron energy from such a model is
compatible with the observations and 
that this model can satisfy the constraints
imposed by observations of the radio emission \cite{kul98}.

	The inferred isotropic energies of GRBs range
up to in excess of $10^{54}$ erg, so if GRB 980425 was in fact
associated with SN1998bw, it was a very puny burst.
Can shock acceleration models account for these much more
luminous cosmological bursts?  To address this question,
Tan et al considered an extreme hypernova model in which
$\ein=5\times 10^{54}$ erg is released into about
$5 M_\odot$ of ejecta.  Whether such an energetic
explosion is possible is not known; whereas most of the energy
in a conventional core-collapse supernova is released in the form
of neutrinos and possibly gravitational radiation, much of
the energy in a hypernova is assumed to go into kinetic form
as the result of a rapidly rotating, magnetized
black hole at the center \cite{pac98}.  Tan et al find that
in this model almost 1\% of the explosion energy goes into material
with $\Gamma_f\beta_f>10$.  The principal observational limits on the
minimum value of $\Gamma_f\beta_f$ come from the constraints
that the opacity due to 
photon-photon interactions 
and to the external medium be small.
These limits in turn depend upon the size of the burst and
therefore the timescale of the variations.  While some GRBs show
very rapid variability, most of the bursts with observed afterglows
(and therefore measured energies) are relatively smooth.
By assuming that the variations observed in these bursts
could be explained by fluctuations in the ambient medium
or by instabilities in the ejecta, Tan et al inferred that
the minimum values of $\Gamma_f\beta_f$ were of order
10, almost an order of magnitude smaller than the limits
inferred by some other workers (see \cite{pir99}).  As 
a result, they concluded that some of the cosmological GRBs
with measured afterglows could be accounted for by an
extreme hypernova model with a spherical explosion.  However,
a number of the bursts were too energetic to be accounted for
by such a model.  The most likely conclusion is that the
cosmological bursts are aspherical, so that the 
emission is beamed towards us and the actual explosion energy
is much smaller than the isotropic value (see \cite{pir99}).
Furthermore, beaming provides the most natural explanation for the 
optical and X-ray
light curves of some bursts, such as GRB 990510
\cite{sta99} \cite{pia00}.

	Tan et al conclude that very young SNRs (including very
young hypernova remnants) might indeed be the engines that underlie
GRBs.  Whereas the explosion that produced the GRB that is
believed to be associated with SN1998bw could have been
approximately spherical, the explosions associated with the much
more energetic cosmological bursts are most likely quite
aspherical.
The recently launched {\it HETE-2} spacecraft,
the proposed {\it Swift} mission, and the AASC-recommended
{\it Gamma-ray Large Area Space Telescope (GLAST)} and {\it EXIST}
missions should provide the data that will enable us to
finally determine the nature of the enigmatic GRBs.

\section*{Conclusion: The Connections Challenge}

	In this brief review, I have tried to show how YSNRs
are at the nexus of many important problems in contemporary
astrophysics: the late stages of stellar evolution, supernovae,
the formation of compact objects, nucleosynthesis,
the formation and destruction of interstellar dust, astrophysical
gas dyanamics, and possibly even gamma-ray bursts.  In order for
the study of YSNRs to reach its potential, however, we must
overcome a major challenge, which I call the ``{\it Connections
Challenge}:''  How can one infer the nature of the supernova
and its progenitor from observations of the YSNR?  Some steps
have been taken in this direction for the best-studied YSNRs 
such as Cas A (e.g., \cite{fes88}), but much more needs to be done.
A key step in this process is to identify the compact object
that remains from the explosion, when it exists.  
In some cases this can be done through observations
of radio pulsars.  However, the recent discovery of an
X-ray point source in Cas A \cite{tan99} shows that other
approaches to this problem are possible and opens up a new
window on understanding supernovae and the formation
of compact objects \cite{ume00}.  

	The results presented at this
conference show that we are making progress in addressing
the challenges I have presented.  If the ambitious
program for new instruments and theory recommended
by the AASC is carried out,
the coming decade could well see the resolution
of most of these challenges.

{\it Acknowledgments.}
I wish to thank the organizers for inviting me to an extremely
valuable conference. Comments on a draft of this paper by Roger
Chevalier, Chris Matzner, and
Jonathan Tan are gratefully acknowledged.  My research is supported in part by 
the National Science Foundation under grant AST 95-30480.


\begin{references}
\bibitem{chev76} Chevalier, R.A. 1976, ApJ, 207, 872
\bibitem{chev89} Chevalier, R.A., \& Soker, N. 1989, ApJ, 341, 867
\bibitem{mm99}   Matzner, C.D., \& McKee, C.F. 1999, ApJ, 510, 379
\bibitem{tan00}  Tan, J.C., Matzner, C.D., \& McKee, C.F. 2000, ApJ
		 (in press), astro-ph 0012003
\bibitem{dwar98} Dwarkadas, V.V., \& Chevalier, R.A. 1998, ApJ, 497, 807
\bibitem{mck74}  McKee, C. F. 1974, ApJ, 188, 335
\bibitem{ham84}  Hamilton, A.J.S., \& Sarazin, C.L. 1984, ApJ, 281, 682
\bibitem{chev82} Chevalier, R.A. 1982, ApJ, 258, 790
\bibitem{nad85}  Nadyozhin, D.K. 1985, Ap\&SS, 112, 225
\bibitem{mck95}  McKee, C.F., \& Truelove, J.K. 1995, Phys. Rep., 256, 157
\bibitem{true99} Truelove, J.K., \& McKee, C.F. 1999, ApJS, 120, 299
\bibitem{true00} Truelove, J.K., \& McKee, C.F. 2000, ApJS, 128, 403
\bibitem{hug00}  Hughes, J.P. 2000, ApJ (submitted), astro-ph 0010122
\bibitem{chev98} Chevalier, R.A. 1998, Memorie della Societa
		 Astronomia Italiana, 69, 977 
\bibitem{blon00} Blondin, J.M., Borkowski, K., \& Reynolds, S.P. 2000,
		 ApJ (submitted), astro-ph 0010285
\bibitem{maj98}  Maciejewski, W., \& Cox, D.P. 1998, ApJ, 511, 792 
\bibitem{gull73} Gull, S.F. 1973, MNRAS, 161, 47
\bibitem{chev92} Chevalier, R.A., Blondin, J.M., \& Emmering,
		 R.T. 1992, ApJ, 392, 118
\bibitem{aasc}   McKee, C.F., \& Taylor, J.H. 2001, {\it Astronomy and 
		 Astrophysics in the New Millennium}, National Academy
		 Press (in press).   
\bibitem{clay75} Clayton, D.D. 1975, ApJ, 199, 765
\bibitem{dwek98} Dwek, E. 1998, ApJ, 501, 643
\bibitem{ds79}   Draine, B.T., \& Salpeter, E.E. 1979, ApJ, 231, 438
\bibitem{mck89}  McKee, C.F. 1989, in {\it IAU Symp. 135 Interstellar
		 Dust}, ed. L.J. Allamandola \& A.G.G.M. Tielens,
		 Dordrecht: Kluwer, 431
\bibitem{tiel94} Tielens, A.G.G.M., McKee, C.F., Seab, C.G., and
		 Hollenbach, D.J., 1994, ApJ, 431, 321
\bibitem{dwek96} Dwek, E., Foster, S.M., \& Vancura, O. 1996, ApJ,
		 457, 244
\bibitem{jone96} Jones, A.P., Tielens, A.G.G.M., \& Hollenbach,
		 D.J. 1996, ApJ, 469, 740
\bibitem{douv99} Douvion, T., Lagage, P.O., \& Cesarsky, C.J. 1999,
		 A\&A, 352, L111
\bibitem{aren99} Arendt, R.G., Dwek, E., \& Mosely, S.H. 1999, ApJ,
		 521, 234
\bibitem{hug00b} Hughes, J.P., Rakowski, C.E., \& Decourchelle,
		 A. 2000, ApJ, 543, L61 
\bibitem{ghav01} Ghavamian, P., Raymond, J., Hartigan, P., \& Blair,
		 W.P. 2001, ApJ (submitted), astro-ph 0010496
\bibitem{mck75}  McKee, C.F. \& Cowie, L.L. 1975, ApJ, 195, 715
\bibitem{mck88}  McKee, C.F. 1988, in {\it IAU Colloq. 101, Supernova
		 Remnants and the Interstellar Medium}, ed. R.S. Roger
		 and T.L. Landecker, Cambridge: Cambridge University
		 Press, p. 205
\bibitem{kle94}  Klein, R.I., McKee, C.F., \& Colella, P. 1994, ApJ,
		 420, 213
\bibitem{kle00}  Klein, R.I., Budil, K.S., Perry, T.S., \& Bach, D.R.
		 2000, ApJS, 127, 379
\bibitem{mck77}  McKee, C.F., \& Ostriker, J.P. 1977, ApJ, 218, 148
\bibitem{beg90}  Begelman, M.C., \& Fabian, A.C. 1990, MNRAS, 244, 26P 
\bibitem{sla93}  Slavin, J.D., Shull, J.M., \& Begelman, M.C. 1993,
		 ApJ, 407, 83
\bibitem{fes96}  Fesen, R.A., \& Gunderson, K.S. 1996, ApJ, 470, 967
\bibitem{sen01}  Senda, A., et al 2001, poster 4.03 at this conference
\bibitem{van91}  van den Bergh, S., \& Tammann, G.A. 1991, {\it Annual
		 Reviews of Astronomy and Astrophysics}, 29, 363
\bibitem{teg00}  Tegmark, M., Zaldarriaga, M., \& Hamilton,
		 A.J.S. 2000, astro-ph 0008167
\bibitem{van94}  van den Bergh, S., \& McClure, R.D. 1994, ApJ, 425, 205
\bibitem{mck97}  McKee, C.F., \& Williams, J.P. 1997, ApJ, 476, 144
\bibitem{sca86}  Scalo, J. 1986, Fund. Cosmic Phys., 11, 1
\bibitem{pir99}  Piran, T. 1999, Phys. Reports, 314, 575
\bibitem{kul00}  Kulkarni, S.R., et al 2000, in {\it Gamma-Ray Bursts: 5th
		 Huntsville Symposium}, eds, Kippen, R.M., Malozzi, R.S.,
		 \& Fishman, G.J., New York, AIP, p. 277
\bibitem{gal98}  Galama, T.J., et al 1998, Nature, 395, 670
\bibitem{col74}  Colgate, S.A. 1974, ApJ, 187, 333
\bibitem{pac98}  Paczynski, B.P. 1998, in {\it Gamma-Ray Bursts: 4th
		 Huntsville Symposium}, eds, C.A. Meegan, R.D. Preece,
		 \& T.M. Koshut, New York, AIP
\bibitem{woo99}  Woosley, S.E., Eastman, R.G., \& Schmidt, B.P. 1999,
		 ApJ, 516, 788
\bibitem{gna85}  Gnatyk, B.I. 1985, Sov. Astron. Lett., 11(5), 331
\bibitem{joh71}  Johnson, M.H., \& McKee, C.F. 1971, PRD, 3, 858
\bibitem{kul98}  Kulkarni, S.R., et al 1998, Nature, 395, 663
\bibitem{sta99}  Stanek, K.Z., et al. 1999, ApJ, 522, L39
\bibitem{pia00}  Pian, E., et al 2000, A\&A (submitted), astro-ph 0012107
\bibitem{fes88}  Fesen, R.A., Becker, R.A., \& Goodrich, R.W. 1988,
		 ApJ, 329, L89	
\bibitem{tan99}  Tananbaum, H. 1999, IAU Circular 7246
\bibitem{ume00}  Umeda, H., Nomoto, K., Tsuruta, S., \& Mineshige,
		 S. 2000, ApJ, 534, L193
 


\end{references}
\end{document}